# Landau's Major Contribution to Physics: A Brief Overview

C Sivaram and Kenath Arun

Indian Institute of Astrophysics, Bangalore

Lev Davidovich Landau, one of the world's greatest and most versatile physicists, was born in Baku, Azerbaijan, Russia on January 22, 1908. He studied physics at Baku University and in Leningrad and graduated in 1927. From 1929-31, he visited various centres in Europe, including Gottingen, the Bohr Institute at Copenhagen and then Cambridge University to work with Paul Dirac.

On his return in 1932, he became head of the theoretical division at the Ukrainian Institute at Kharkov, establishing the first school of theoretical physics in the former Soviet Union, which soon received international reputation. In 1937, the eminent Russian physicist, Peter Kapitza who was then the director of the Institute for Physical Problems in Moscow, persuaded Landau to head the theory division at the Institute.

Landau's work spans a very wide range and has had a considerable impact on all areas of physics including condensed matter physics, plasma, high energy and particle physics, fluid dynamics, astrophysics, gravitation, elasticity, etc.

Along with his former fellow student, E Lifshitz, he wrote the set of a dozen volumes, [1] the magnificent world-renowned 'Course of Theoretical Physics' which covered topics ranging from mechanics, classical theory of fields, quantum mechanics, electrodynamics of continuous media, fluid dynamics, kinetic theory, theory of elasticity, statistical physics and more.

"There is hardly anything in physics which the authors have not mastered!"

-Nature

Some well known results of Landau's work include:[1, 2]

1. Theory of liquid helium. Introduction of notion of quasiparticles. Superfluid helium as a quantum liquid. Phonons and rotons. Prediction of second, third, fourth and fifth sounds, zero sound.
2. Ginzburg-Landau theory of superconductivity. Type II superconductors, maximum magnetic field, led to concept of magnetic vortices carrying quantised flux.
3. Landau damping in plasma
4. Landau-Lifshitz equations for evolution of spin fields in ferromagnets. General theory of phase transitions.
5. Landau levels in diamagnetism
6. Elasticity: Attenuation of shear waves in crystals, Landau-Rumer theory
7. Neutron star mass limit
8. High energy physics: Landau-Pomeranchuk theorem; particle physics: two component theory of neutrinos, CP violation, Landau gauge, Landau poles
9. Gravitation: Landau-Lifshitz pseudo-tensor for calculating energy of the gravitational field
10. Hydrodynamics: Theory of explosion, Landau-Darrieus instability (now applied to type Ia supernova)

## 1. Theory of liquid helium

Landau thought of superfluid helium-4 as a quantum liquid so that its atoms behave as if they were in the same quantum state. Their motions are highly correlated enhancing the physical properties of the superfluid state. He introduced the notion of quasiparticles, which act as carriers of motion in the liquid. This is an important concept applied in many other phenomena in condensed matter physics. (For instance, polarons in a crystal is electron 'dressed' by the surrounding lattice deformation).

The energy-momentum (dispersion) relation for elementary excitations in superfluid helium-4 has the following form:

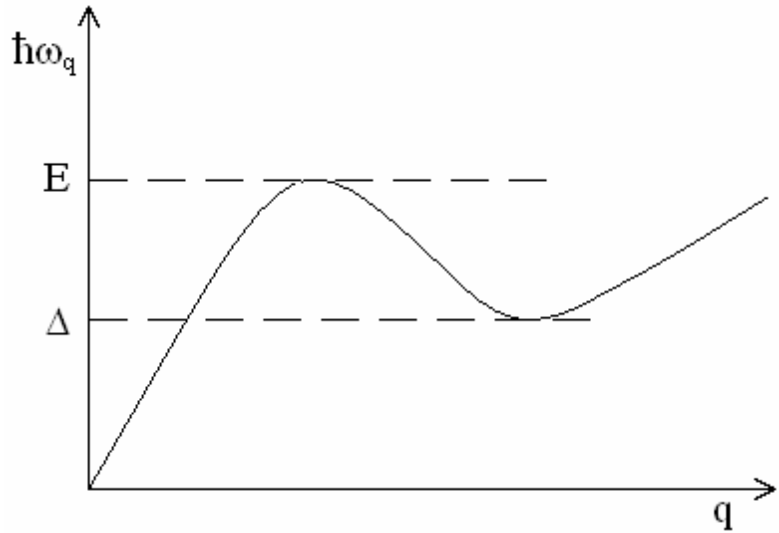

The long wavelength excitations are phonons with dispersion $\hbar\omega_q = c_S q$, where $c_S$ is the sound speed and q is the magnitude of the wave vector. These are the phonons.

In case of shorter wavelengths, the excitation spectrum has a local minimum. The excitations near this minimum are known as rotons and their dispersion relation is roughly described by:

$\hbar\omega_q \approx \Delta + \alpha(q - q_0)^2$, $\alpha$ is a constant.

Density of states $\rho(E) \propto q^2 \left|\dfrac{dq}{dt}\right|$

For wavevector near (minimum) $q_0$, $E \approx \Delta + \alpha k^2$, where, $k = q - q_0$. Defining $m = \dfrac{\hbar^2}{2\alpha}$, by analogy with the non-relativistic relation $E = \dfrac{\hbar^2 k^2}{2\alpha}$, we see that this describes free quasiparticles of mass $m$ in a constant potential $\Delta$.

One can write a single roton wave function (using the NR Schrodinger equation). The binding of rotons is related to phenomenon of Cooper pairs in superconductors. Landau first predicted existence of a second sound in the superfluid. Unlike the usual first sound (which propagates pressure oscillations) second sound involves oscillations in entropy and is a temperature wave in the bulk superfluid.

This was observed experimentally in 1944, in Moscow, confirming many aspects of the two fluid model of liquid helium. The second sound velocity $c_{2S}$ involves the ratio of the superfluid to normal component $\rho_n$ densities, that is $\rho_S / \rho_n$, and is given by:

$$c_{2S}^2 = \dfrac{\rho_S}{\rho_n} \dfrac{TS^2}{C}$$

Where T is the temperature, S is the specific entropy and C is the specific heat.

Above the so called lambda $(\lambda)$ point, when there is no superfluid, $\dfrac{\rho_S}{\rho} = 0$, so $c_{2S} = 0$.

Below about $1K$, there is a rapid increase to value $c_{2S} = c_{1S}/\sqrt{3}$, where $c_{1S}$ is the usual first sound velocity having a value $c_{1S}^2 = \dfrac{B}{\rho}$, B is the bulk modulus and $\rho$ the total density.

The third sound occurs in very thin films and is given by:

$$c_{3S}^2 = \dfrac{\rho_S}{\rho} \dfrac{3\beta}{d^3}$$

Where d is the film thickness, $\beta$ is the Van der Waal coefficient. Typically it is 140m/s.

The fourth sound is a pressure wave propagating in superfluid helium confined to a porous material (like packed powders) and is given by:

$$c_{4S}^2 = \frac{\rho_S}{\rho} c_{1S}^2$$

As $\frac{\rho_S}{\rho} \to 1$, $c_{4S} \to c_{1S}$

This measurement (of $c_{4S}$) can be used to give precise values of the superfluid fraction $\frac{\rho_S}{\rho}$ as a function of T. we also have the fifth sound which is a temperature wave which can propagate in helium confined to a narrow orifice.

It is analogous to the second sound except that only a superfluid component can flow and is a low velocity mode with a maximum of 12m/s at about $2K$ and vanishes at the lambda point where $\rho_n = 0$.

$$c_{5S}^2 = \frac{\rho_n}{\rho} c_{2S}^2$$

Again when the sound frequency is increased to a value greater than the collision rate between helium atoms the wave changes to zero sound from first sound as first predicted by Landau (for He-3) and has quite recently been discovered. (For an account of the above phenomena, one can refer for example to Superfluid Hydrodynamics by S Putterman (1974))

## 2. Ginzburg-Landau theory of superconductivity

Another famous contribution to condensed matter physics is the Ginzburg-Landau theory of superconductivity which is a mathematical way of describing properties of superconductors with the help of general thermodynamic arguments. It is especially useful for understanding Type II superconductors which can be thought of as a second order transition from normal to superconducting phase in a uniform magnetic field.

The phase boundary line for the penetrating magnetic field $H_{C_2}(T)$ as a function of temperature can be calculated using the Ginzburg-Landau equation for the superconducting order parameter $\psi(x)$, valid at all temperatures $T < T_C$ (so called field transition temperature).

The equation is

$$\left(\Delta - i\frac{ze}{\hbar c}A\right)^2 \psi + \lambda^{-2}\psi + \beta|\psi|^2\psi = 0$$

Where $\lambda(T) = \lambda_0\left(1 - T/T_C\right)^{-1/2}$ is the coherence length, $A(x)$ is the vector potential. At the onset of superconductivity, number density of Cooper pairs, $n_s \propto \psi^*\psi$, is small, so the last term in the Ginzburg-Landau equation (so called cubic term in $\psi(x)$, which is a novel feature of the model, the relativistic version leading later to the so called Higgs model with its famous still sought after particle!) can be neglected.

So we have domination by the external field with the substitution $\psi(x) = \varphi(x)e^{ik_y y}e^{ik_z z}$, the equation can be shown to reduce to:

$$\left(-\frac{\hbar^2}{2m}\frac{d^2}{dx^2} + \frac{1}{2}\omega^2 x^2\right)\phi = E\phi$$, the equation for a SHO, with $\omega = 2eB/cm$.

Given $\lambda$, the maximum allowed value of the magnetic field $B_{max}$ (corresponding to $n=0, k_z = 0$) is

$$H_{C^2} = B_{max} = \left(\frac{\hbar c}{2e}\right)\frac{1}{\lambda^2} = \frac{\hbar c}{2e\lambda_0^2}\left(1 - \frac{T}{T_C}\right).$$

When the field exceeds $B_{max}$, sample is no longer superconducting. The field is confined to vortices (the theory was later refined by Abrikosov), the radius of the vortex being $\lambda$, the quantum of flux being $\frac{\hbar c}{2e}$ (in fact the factor of 2 was already noted before Cooper pairs were postulated). So $B_{max} = \frac{\phi}{\pi\lambda^2} = \frac{\hbar c}{2e\pi\lambda^2}$.

This has also been applied to neutron star magnetic fields (with pion condensation transition for example which gives a maximal interior field of $10^{16} G$). The Ginzburg-Landau theory is still of great relevance in understanding the new high temperature superconductors.

Following the tradition of Landau being awarded the Nobel Prize in 1962, for his theory of liquid helium, Ginzburg and Abrikosov shared the 2003 Nobel Prize with Anthony Legett, for his detailed theoretical explanation of He-3 superfluidity.

Also we note above that the electron motion in the z-direction given rise to quantized energy levels $\left(n+\frac{1}{2}\right)\frac{eB}{mc}$, the so called ubiquitous Landau levels.

A recent application has been to estimate neutron star magnetic fields from cyclotron X-ray lines seen in the compact stars, $\frac{eB}{mc}$ the gyro-frequency being in kilo-electron volts for such large magnetic fields.

### 3. Landau damping in plasma

In plasma physics, we have the famous Landau damping[3, 4] (a damping mechanism by which plasma particles absorb wave energy even in a collisionless plasma) which was experimentally discovered in 1965 (by Malemberg and Whorton)[5] 20 years after Landau's prediction.

If the number of particles slightly slower than the phase velocity of the wave is larger than the number slightly faster, that is, if:
$v_0 \frac{\partial f_0}{\partial v_0} < 0$, the group of particles as a whole gains energy from the wave and the wave is damped.

On the contrary when $v_0 \frac{\partial f_0}{\partial v_0} > 0$, at $v_0 = \omega/k$, the particles give their energy to the wave and the wave amplitude increases. This showed that energy exchange processes are possible even in collisionless plasma. This phenomenon has found wide applications ranging from quark-gluon plasma to synchronous light flashing of fireflies!

## 4. Landau-Lifshitz equations for evolution of spin fields in ferromagnets

Again we have the famous Landau-Lifshitz equations to describe the evolution of spin fields in continuum ferromagnets. They play a basic role in understanding non-equilibrium magnetism analogous to that if Navier-Stokes equation in fluid dynamics.

The equation has the form:
$$\frac{\partial u}{\partial t} = \lambda_1 u \times H^{ext} - \lambda_2 u \times (u \times H^{ext})$$

For a zero external filed, we have the dissipation (Gilbert term), $\lambda u \times (u \times \Delta u)$.

There are several interesting solutions to the above equations in various contexts. The general theory of phase transitions was also developed (analogous to the Ginzburg-Landau theory).

## 5. Landau levels in diamagnetism

Landau also made an early contribution to diamagnetism. He calculated the diamagnetism of free electrons (vanishing classically by the Bohr-van Leeuwen theorem) quantum mechanically and for electrons in metals, the numerical value of Landau diamagnetism is 1/3[rd] that of the Pauli spin paramagnetism (for bound electrons there is cancellation of diamagnetic and paramagnetic susceptibilities).

## 6. Landau-Rumer theory

In elasticity we have the Landau-Rumer theory which deals with the direct interaction of acoustic waves with thermal phonons. They showed that attenuation for propagating slow shear waves increases as the fourth power of the temperature.

The attenuation coefficient was proportional $\frac{kT}{M v^2}\left(\frac{T}{\theta}\right)^3 \frac{1}{\lambda}$, where $\lambda$ is the wavelength, $\theta$ the Debye temperature and v the sound velocity.

Experiments were in good agreement with the theory.

## 7. Neutron star mass limit

Landau also obtained a mass limit[6] for neutron stars which was the same as the Chandrasekhar limit. It was given as $M \sim \left(\frac{\hbar c}{G m_n^2}\right)^{3/2} m_n = 1.5 M_{sun}$. Indeed many pulsars (including the Hulse Taylor binary) have just this mass!

It is said that Landau got this result in 1932, very soon after he heard of Chadwick's discovery of the neutron. He realised that this would be the 'ultimate' stellar core! (Matter only made of neutrons). Many Russian articles call this the Landau-Chandrasekhar limit.

## 8. High energy physics

In high energy physics, we have the Landau-Pomeranchuk theorem, which states that the cross-sections for particle and antiparticle interactions should (with increasing energy) asymptotically approach the same value. After the discovery of parity violation in weak interactions, Landau (and independently Lee and Yang and Salam) proposed the two component neutrino theory, where the Weyl equation (rather than the Dirac equation) was the appropriate one for neutrinos.

Landau also elegantly showed that an electric dipole moment for an elementary particle can exist only in the case when there is a violation of not only parity conservation (P) but also time reversal (T). This can be seen by defining electric dipole moment of a particle as follows:

Let $\sigma_J$ be the charge density inside a particle with angular momentum $\bar{J}$, with quantum number $J$, whose orientation is given by $m = J$ relative to the z-axis through the centre of mass. Then

$ed = \int \sigma_J dVz$, $dV$ is the volume element.

If particle is charged then this implies charge centroid does not coincides with mass centroid when $d \neq 0$. For a particle like a neutron (uncharged), then $d \neq 0$ corresponds to excess positive or negative charge in one hemisphere. To connect with CP, we see that if $d = 0$ there is symmetry under $P$ and $T$ ($d$ does not change under $T$, but $J$ does, so $d$ must vanish if there is $T$ symmetry). So if CPT is a good symmetry, $T$ violation implies CP violation. Thus $d \neq 0$ only if CP is violated.

Landau pole occurs in the evolution of the (electromagnetic) coupling constant at a particular value of the momentum scale when perturbation theory breaks down and the coupling diverges.
Thus:

$$\bar{e}^2(p) = \frac{e^2}{1 - 2\beta e^2 p} = \frac{1}{\frac{1}{e^2} - 2\beta_0 p}$$

As $p$ increases, $e^2$ grows and diverges when $p \to \frac{1}{2\beta_0 e^2}$

For QCD, the behaviour is opposite, the coupling vanishes as the momentum tends to infinity.

$$g^2(t) = \frac{g^2}{1 - 2|\beta_0|g^2 p} = \frac{1}{\frac{1}{g^2} + 2|\beta_0|p}, \quad (p \to \infty, g \to 0)$$

Landau gauge: For the gauge fixing term

$L_{GF} = \frac{\lambda}{2}\phi^a\phi^a + (\partial^\mu\phi^a)A_\mu^a$; ($\phi^a$ is the auxiliary field, $\lambda$ is the gauge fixing parameter)

$\lambda = 0$, corresponds to the Landau gauge.

Again for the propagator

$$D_{\mu\nu}^{ab}(p) = -\frac{i\delta^{ab}}{p^2}\left(\eta_{\mu\nu} + (\lambda-1)\frac{p_\mu p_\nu}{p^2}\right),$$

$\lambda = 0$ is the Landau gauge, when propagator is transverse

$\lambda = 1$ is the Feynman gauge, which makes the form simple, that is: $D_{\mu\nu}^{ab}(p) = -\frac{i\delta^{ab}}{p^2}\eta_{\mu\nu}$

## 9. Landau-Lifshitz pseudo-tensor

A common prescription for the energy momentum tensor describing the energy carried by gravitational waves is the Landau-Lifshitz pseudo-tensor

$$T_{\mu\nu}^{G\omega} = \frac{c^4}{32\pi G}\left\langle \bar{h}_{\alpha\beta,\mu}h_{,\nu}^{\alpha\beta} - \frac{1}{2}\bar{h}_{,\mu}\bar{h}_{,\nu}\right\rangle$$

In the TT gauge (transverse-traceless gauge, where there are only two components) the second term in the angle brackets vanishes. This is the analogue of the Poynting vector describing the energy carried by electromagnetic waves.[7]

The energy flux density of the wave is given as:

$$T_{01} = \frac{1}{32\pi G}\left\langle \bar{h}_{\alpha\beta,0}\bar{h}_{,1}^{\alpha\beta}\right\rangle = \frac{\omega^2}{32\pi G}\times 2h^2 \times \frac{1}{2} = \frac{\omega^2}{32\pi G}h^2$$

Factor ½ comes from taking time average of square of an oscillating quantity, h is the wave amplitude, $\omega$ the frequency.

If $T_{01} = \frac{c^5/G}{4\pi r^2}$ (maximal flux), then at $\omega = 1kHz$, $r = 1kpc$, we have $h \sim 10^{-14}$, $r = 1Mpc$, we have $h \sim 10^{-17}$.